\documentclass[runningheads,fleqn]{svmult}

\usepackage{graphicx}  
\usepackage{physproc}  

\newcommand{\greeksym}[1]{{\usefont{U}{psy}{m}{n}#1}}

\newcommand{\uDelta}{\mbox{\greeksym{D}}}

\renewcommand{\Delta}{\uDelta}

\newcommand{\figurewidth}{0.9\textwidth}

\begin{document}

\title*{Geometric Cluster Algorithm for\newline Interacting Fluids}

\toctitle{Geometric Cluster Algorithm for Interacting Fluids}

\titlerunning{Geometric Cluster Algorithm for Interacting Fluids}

\author{Erik Luijten \and Jiwen Liu}

\authorrunning{Erik Luijten and Jiwen Liu}

\institute{University of Illinois at Urbana-Champaign, Urbana, IL 61801, USA}

\maketitle

\begin{abstract}
  We discuss a new Monte Carlo algorithm for the simulation of complex fluids.
  This algorithm employs geometric operations to identify clusters of particles
  that can be moved in a rejection-free way. It is demonstrated that this
  \emph{geometric cluster algorithm}~(GCA) constitutes the continuum
  generalization of the Swendsen--Wang and Wolff cluster algorithms for spin
  systems. Because of its nonlocal nature, it is particularly well suited for
  the simulation of fluid systems containing particles of widely varying sizes.
  The efficiency improvement with respect to conventional simulation algorithms
  is a rapidly growing function of the size asymmetry between the constituents
  of the system. We study the cluster-size distribution for a Lennard-Jones
  fluid as a function of density and temperature and provide a comparison
  between the generalized GCA and the hard-core GCA for a size-asymmetric
  mixture with Yukawa-type couplings.
\end{abstract}

\section{Introduction and Motivation}

The presence of multiple time and length scales constitutes one of the major
problems in computer simulations of matter. In simulations that faithfully
capture the dynamic evolution of a system, the fastest particles in the system
dictate the required time resolution. If different types of particles with
widely varying diffusion rates are present, then the slower particles may be
unable to explore the entire configuration space during the course of the
simulation, leading to ergodicity problems.

In the computational study of complex fluids, such as colloidal suspensions,
this problem frequently occurs, since such systems typically contain particles
of different sizes.  A concrete example is the `nanoparticle haloing'
phenomenon discovered by Lewis and coworkers~\cite{tohver01}. This experimental
work deals with a new approach to the stabilization of suspensions of
micron-sized spherical particles, which tend to aggregate under the influence
of their mutual van der Waals attraction. Conventional approaches to prevent
this gelation, such as charge stabilization (variation of the pH to alter the
surface charge of the particles), lead to complications in certain
applications, e.g., in the formation of colloidal crystals, where the
electrostatic repulsion prevents the close packing of the particles.  It was
found that these complications can be avoided by generating an effective
colloidal repulsion through the addition of highly-charged nanometer-sized
particles to the suspension of the (near-neutral) colloids. A concrete
explanation for the underlying mechanism leading to the effective repulsions is
currently lacking. Evidently, a computational approach to this problem must
involve both the microspheres and the nanoparticles, which typically differ by
a factor~100 in diameter. Traditionally, the effective pair interaction between
colloids (potential of mean force) is then calculated by ``integrating out''
the smaller species, e.g., by simulating a system containing two spheres at
fixed separation embedded in a sea of smaller
particles~\cite{dickman97,allahyarov98}. Here, we discuss a new simulation
algorithm~\cite{geomc} that is capable of explicitly incorporating both species
and computing equilibrium properties of the suspension, without suffering from
the disparity in time scales. This algorithm can be viewed as a continuum
version of the widely-used cluster algorithms for lattice spin
models~\cite{swendsen87,wolff89}.

\section{Cluster Monte Carlo Algorithms}

Equilibrium Monte Carlo methods are aimed at obtaining static thermodynamic and
structural properties by generating system configurations according to the
Boltzmann distribution. Nonphysical moves may be employed in the underlying
Markov process, which -- in principle -- offers an exquisite way to overcome
the presence of multiple time scales. Thus, Monte Carlo algorithms are the
method of choice for, e.g., simulations of critical phenomena and phase
transitions. In the context of size-asymmetric fluids, collective moves have
been devised in which groups of particles are moved simultaneously. Unless
these groups are identified in a careful way, which typically involves
knowledge about the physical properties of the system, the Monte Carlo step
will entail a large energy change and consequently have only a very small
acceptance rate. Thus, while this approach has been used quite successfully in
specific situations~\cite{wu92,jaster99,woodcock99,lobaskin99}, it typically
involves a tunable parameter and cannot be generalized in a straightforward
fashion.

The situation is rather different for spin models, for which Swendsen and Wang
(SW)~\cite{swendsen87} introduced a \emph{cluster algorithm} in which groups of
parallel spins are identified in a probabilistic manner, based upon the
Fortuin--Kasteleyn mapping of the Potts model onto the random-cluster
model~\cite{fortuin72}. These groups can subsequently be flipped independently.
This rejection-free algorithm (every completed cluster can be flipped without
an additional evaluation of the resulting energy difference) suppresses
critical slowing down. Accordingly, its generalization to off-lattice fluids
has been a widely-pursued goal. However, the SW algorithm -- as well as the
even more efficient single-cluster variant introduced by Wolff~\cite{wolff89}
-- relies on invariance of the Hamiltonian under a spin-inversion operation.
For a fluid, this translates into particle--hole symmetry, which is only obeyed
for lattice gases. Consequently, efficient off-lattice cluster algorithms have
only been designed for a small number of specific fluid
models~\cite{johnson97,sun00}, and it is not clear how these methods can be
generalized.  Hard-sphere fluids constitute another special case. Here every
configuration without particle overlaps has the same energy. Dress and
Krauth~\cite{dress95} devised a strategy to create such configurations by means
of geometric transformations. Their approach is particularly advantageous in
size-asymmetric mixtures, but cannot be applied to systems with other
interactions without supplementing it with a costly acceptance criterion.
Nevertheless, a remarkable feature of this geometric cluster algorithm is that
it relies on the invariance of the Hamiltonian under geometric operations.
Here, we exploit this property to formulate a cluster Monte Carlo algorithm
that is applicable to arbitrary pair potentials without the imposition of an
acceptance criterion.

\section{Generalized Geometric Cluster Algorithm}

\begin{figure*}[b]
\begin{center}
\includegraphics[width=\textwidth]{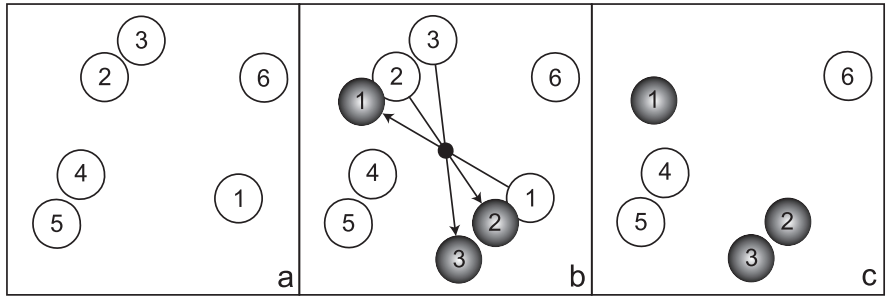}
\caption{Two-dimensional illustration of the interacting geometric
cluster algorithm. Open and shaded circles label the particles before and after
the geometrical operation, respectively. The small filled disk denotes the
pivot.  a)~Initial configuration; b)~construction of a new cluster via point
reflection of particles $1$--$3$ with respect to the pivot; c)~final
configuration.}
\label{fig:geomc}
\end{center}
\end{figure*}

\subsection{Single-cluster variant}

The geometric cluster algorithm as formulated by Dress and Krauth starts from a
configuration of particles, with periodic boundary conditions. This
configuration is rotated around an arbitrarily chosen pivot.  Groups of
particles that overlap between the original and the rotated configuration are
exchanged between these configurations.  In practice, it is more convenient to
construct only a single cluster and to carry out a point reflection on the
fly~\cite{heringa98}. For each cluster a new pivot is chosen.  This constitutes
the counterpart of the Wolff algorithm for spin models~\cite{wolff89}. In the
presence of a general, isotropic pair potential~$V(r)$ the cluster construction
proceeds as follows (cf.  Fig.~\ref{fig:geomc}):
\begin{enumerate}
\item Choose a random pivot point.
\item Choose the first particle~$i$ at random and move it from its original
  position~$\mathbf{r}_i$ to its new position~$\mathbf{r}_i'$, via a point
  reflection with respect to the pivot.
\item Identify all particles that interact with~$i$ in its original position or
  in its new position. These particles~$j$ are considered for a point
  reflection with respect to the pivot. This reflection is carried out with a
  probability $p_{ij} = \max[1 - \exp(-\beta \Delta_{ij}), 0]$, where
  $\Delta_{ij} = V(|\mathbf{r}_i'-\mathbf{r}_j|) -
  V(|\mathbf{r}_i-\mathbf{r}_j|)$ and $\beta = 1/k_{\rm B}T$. Note that
  $p_{ij}$ solely depends on the interaction strength between $i$ and~$j$.
\item Repeat step~3 in an iterative fashion for each particle~$j$ that is added
  to the cluster. If $j$ is moved with respect to the pivot, then all
  its interacting neighbors that have not yet been added to the cluster are
  considered for inclusion as well. The cluster construction is completed once
  all interacting neighbors have been considered.
\end{enumerate}

It is instructive to compare this prescription to the Wolff cluster algorithm,
in which a cluster of parallel spins is grown from a randomly chosen initial
spin. Parallel spins with a ferromagnetic coupling constant~$K$ are added to
the cluster with a probability $1-\exp(-2K)$. The energy difference between the
parallel and the antiparallel configuration indeed equals~$2K$. An antiparallel
spin is never added to the cluster, in accordance with the fact that such a
pair will be in a state of lower energy if only the first spin is flipped.

This algorithm is ergodic, since each particle can be moved over an arbitrarily
small distance. Namely, there is a non-vanishing probability that a cluster
consists of only a single particle and the pivot can be located arbitrarily
close to the center of this particle. Detailed balance is proven by considering
a configuration~$X$ which is transformed into a new configuration~$Y$ by means
of a cluster move. The energy change $E_Y-E_X$ results from every particle that
is \emph{not} included in the cluster but interacts with one or more particles
that are part of the cluster. Each ``broken bond'' has a probability~$1-p_k$,
so that the total probability of forming a cluster is given by
\begin{equation}
\label{eq:xy}
  T(X\to Y) = C \prod_{k} (1-p_k) \;.
\end{equation}
The prefactor $C$ accounts for the probability of choosing a specific pivot and
for the probability of creating a specific arrangement of bonds inside the
cluster. The total set $\{k\}$ of broken bonds can be divided into two subsets.
The broken bonds~$l$ that lead to an increase $\Delta_l$ in pair energy have a
probability $\exp(-\beta \Delta_{l})$, whereas the broken bonds $m$ that lead
to a decrease in pair energy have a probability equal to unity. Accordingly,
the transition probability can be written as
\begin{equation}
\label{eq:xy-exp}
  T(X\to Y) = C \exp\left[-\beta \sum_{l} \Delta_l\right] \;.
\end{equation}
The reverse move, in which the configuration $Y$ is transformed into the
configuration $X$ by moving a cluster that is constructed in the same way, has
a probability
\begin{equation}
  T(Y\to X) = C \prod_{k} (1-\bar{p}_k) \;,
\end{equation}
where the sum runs over the same set $\{k\}$ of broken bonds as in
Eq.~(\ref{eq:xy}), but the sign of all energy differences $\Delta_k$ has been
reversed (indicated by $\bar{p}_k$). Accordingly, the sum over $\Delta_l$ in
Eq.~(\ref{eq:xy-exp}) is replaced by the negative sum over the complementary
set $\{m\}$,
\begin{equation}
  T(Y\to X) = C \exp\left[+ \beta \sum_{m} \Delta_m\right] \;.
\end{equation}
The point reflection is a self-inverse operation, so that detailed balance is
obeyed without the need to impose an acceptance criterion:
\begin{equation}
\frac{T(X\to Y)}{T(Y\to X)} =
 \exp\left[ -\beta \sum_{k} \Delta_k\right] =
 \frac{\exp(-\beta E_{Y})}{\exp(-\beta E_{X})}
 \;.
\end{equation}
Since energy differences are taken into account on a pair-wise basis during the
construction of the cluster, rather than as a total energy difference after the
cluster has been completed, the cluster is constructed in such a way that large
energy differences are avoided. The clusters are representative of the actual
structure of the system.

\subsection{Multiple-cluster variant}

In order to demonstrate that the generalized geometric cluster algorithm (GCA)
indeed constitutes the off-lattice counterpart of the SW and Wolff cluster
algorithms, it is instructive to formulate a multiple-cluster variant. This
formulation yields a full decomposition of an off-lattice fluid configuration
into \emph{stochastically independent clusters}. In the following, we
demonstrate how it can be phrased as a natural extension of the single-cluster
version.

First, a cluster is constructed according to the Wolff version of the GCA, with
the exception that the cluster is only \emph{identified}; particles belonging
to the cluster are marked but not actually moved.  The chosen pivot will also
be used for the construction of all subsequent clusters in this decomposition.
These subsequent clusters are built just like the first cluster, except that
particles that are already part of an earlier cluster will never be considered
for a new cluster. Once each particle is part of a cluster the decomposition is
completed and each cluster is moved with a probability~$f$.

Although all clusters except the first are built in a \emph{restricted}
fashion, every individual cluster is constructed according to the rules of the
Wolff formulation of the GCA\@. The exclusion of particles that are already
part of another cluster simply reflects the fact that every bond should be
considered only once. If a bond is broken during the construction of an earlier
cluster it must not be re-established during the construction of a subsequent
cluster.

In order to establish that this prescription is a true equivalent of the SW
algorithm, we prove that each cluster can be moved (reflected) independently
while preserving detailed balance. If only a single cluster is actually moved,
this essentially corresponds to the Wolff version of the GCA, since each
cluster is built according to the GCA prescription.  The same holds true if
several clusters are moved and no interactions are present between particles
that belong to different clusters (the hard-sphere algorithm is a particular
realization of this situation). If two or more clusters are moved and broken
bonds exist between these clusters, i.e., a non-vanishing interaction exists
between particles that belong to disparate (moving) clusters, then the shared
broken bonds are actually preserved and the proof of detailed balance provided
in the previous section no longer applies in its original form.  However, since
these bonds are identical in the forward and reverse move, the corresponding
factors cancel out. This is illustrated for the situation of two clusters whose
construction involves, respectively, two sets of broken bonds $\{k_1\}$ and
$\{k_2\}$. Each set comprises bonds $l$ that lead to an \emph{increase} in pair
energy and bonds $m$ that lead to a \emph{decrease} in pair energy. We further
subdivide these sets into \emph{external} bonds that connect cluster 1 or~2
with the remainder of the system and \emph{joint} bonds that connect cluster 1
and~2. Accordingly, the probability of creating cluster~1 is given by
\begin{equation}
  C_1 \prod_{i \in \left\{l_1\right\}}(1-p_i) =
  C_1 \prod_{i \in \left\{l_1^{\rm ext}\right\}} (1-p_i)
      \prod_{j \in \left\{l_1^{\rm joint}\right\}} (1-p_j) \;.
\end{equation}
Upon construction of the first cluster, the creation of the second cluster has
a probability
\begin{equation}
  C_2 \prod_{i \in \left\{l_2^{\rm ext}\right\}} (1-p_i) \;,
\end{equation}
since all joint bonds in $\{l_1^{\rm joint}\}$ already have been broken.  The
factors $C_1$ and $C_2$ account for the probability of realizing a particular
arrangement of internal bonds in clusters 1 and~2, respectively.  Hence, the
total transition probability for moving both clusters (upon fixing the pivot)
is given by
\begin{equation}
  T_{12} = C_1 C_2 \exp\left[
  -\beta \sum_{i \in \left\{l_1^{\rm ext}\right\}} \Delta_i
  -\beta \sum_{j \in \left\{l_2^{\rm ext}\right\}} \Delta_j
  -\beta \sum_{n \in \left\{l_1^{\rm joint}\right\}} \Delta_n
  \right] \;.
\end{equation}
In the reverse move, the energy differences for all external broken
bonds have changed sign, but the energy differences for the joint bonds
connecting cluster 1 and~2 are the same as in the forward move. Thus, cluster~1
is created with probability
\begin{eqnarray}
 && C_1 \prod_{i \in \left\{m_1^{\rm ext}\right\}} (1-\bar{p}_i)
      \prod_{j \in \left\{l_1^{\rm joint}\right\}} (1-p_j) \nonumber \\
 &=&
  C_1 \prod_{i \in \left\{m_1^{\rm ext}\right\}}  \exp[ +\beta \Delta_i ]
      \prod_{j \in \left\{l_1^{\rm joint}\right\}} \exp[-\beta \Delta_j ] \;,
\end{eqnarray}
where the $\bar{p}$ reflects the sign change compared to the forward move and
the product over the external bonds involves the complement of the set
$\{l_1^{\rm ext}\}$. The creation probability for the second cluster is
\begin{equation}
  C_2 \prod_{i \in \left\{m_2^{\rm ext}\right\}} (1-\bar{p}_i) =
  C_2 \prod_{i \in \left\{m_2^{\rm ext}\right\}} \exp[ +\beta \Delta_i ]
\end{equation}
and the total transition probability for the reverse move is
\begin{equation}
  \tilde{T}_{12} = C_1 C_2 \exp\left[
  + \beta \sum_{i \in \left\{m_1^{\rm ext}\right\}} \Delta_i
  + \beta \sum_{j \in \left\{m_2^{\rm ext}\right\}} \Delta_j
  -\beta \sum_{n \in \left\{l_1^{\rm joint}\right\}} \Delta_n
  \right] \;.
\end{equation}
Accordingly, detailed balance is still fulfilled,
\begin{equation}
  \frac{T_{12}}{\tilde{T}_{12}} =
  \exp\left[ 
  -\beta \sum_{i \in \left\{k_1^{\rm ext}\right\}} \Delta_i
  -\beta \sum_{j \in \left\{k_2^{\rm ext}\right\}} \Delta_j
  \right] =
  \exp\left[ - \beta (E_Y - E_X ) \right]  \;,
\end{equation}
in which $\{k_1^{\rm ext}\} = \left\{l_1^{\rm ext}\right\} \cup \left\{m_1^{\rm
  ext}\right\}$ and $\{k_2^{\rm ext}\} = \left\{l_2^{\rm ext}\right\} \cup
\left\{m_2^{\rm ext}\right\}$ and $E_X$ and $E_Y$ refer to the total internal
energy before and after the move, respectively. This treatment applies to any
simultaneous move of clusters, so that each cluster in the decomposition indeed
can be moved independently without violating detailed balance. It is noteworthy
that the probabilities for breaking joint bonds in the forward and reverse
moves cancel only because the probability in the cluster construction
factorizes into pairwise probabilities, as opposed to the probability for a
multiple-particle move in a Metropolis-type algorithm.

\begin{figure}
\begin{center}
\includegraphics[width=\figurewidth]{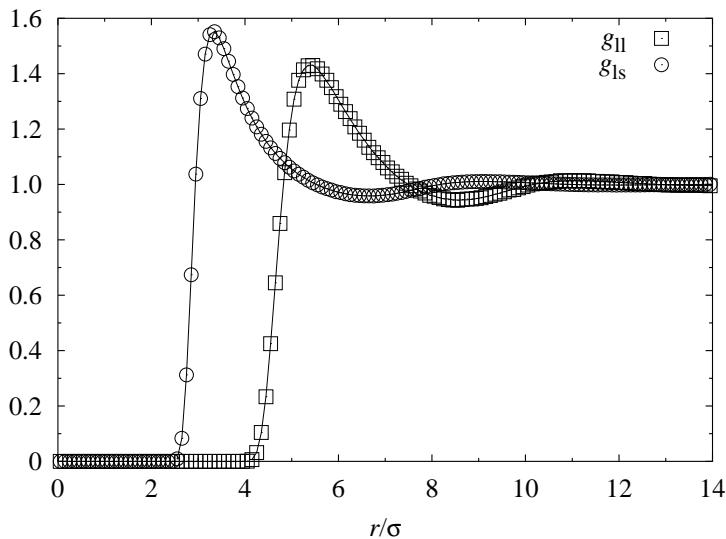}
\caption{Comparison between the single-cluster version (solid lines) and the
multiple-cluster version (symbols) of the generalized geometric cluster
algorithm. The figure shows pair correlation functions for the size-asymmetric
Lennard-Jones mixture described in Ref.~\protect\cite{geomc}. The system
contains $800$ small (diameter $\sigma$) and $400$ large particles (diameter
$5\sigma$) at a total packing fraction $\eta \approx 0.213$. $g_{\textrm{ll}}$
and $g_{\textrm{ls}}$ represent the large--large and large--small correlation
functions, respectively.}
\label{fig:wolff-sw}
\end{center}
\end{figure}

As demonstrated in Fig.~\ref{fig:wolff-sw} for a binary mixture, the results
obtained by means of the multiple-cluster geometric algorithm agree perfectly
with those obtained using the single-cluster version.

\section{Performance}

\begin{figure}[t]
\begin{center}
\includegraphics[width=\figurewidth]{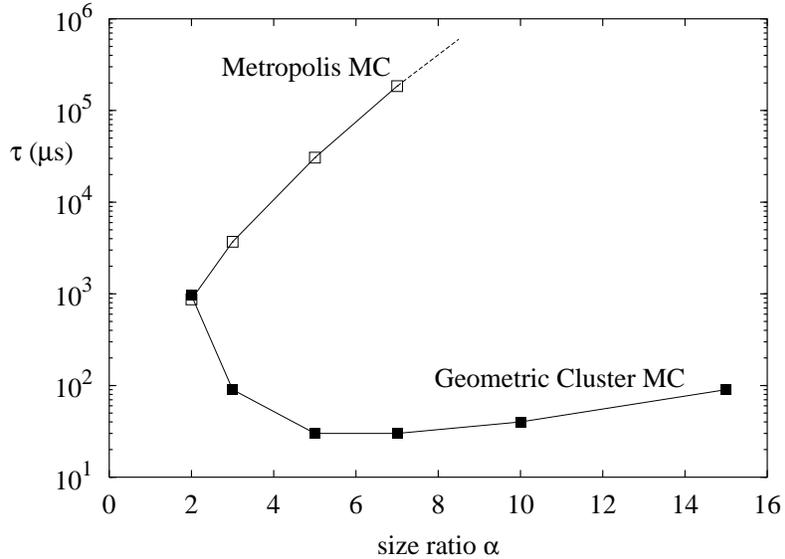}
\caption{Efficiency comparison between a conventional local update algorithm
(open symbols) and the generalized geometric cluster algorithm (closed
symbols), for a binary mixture (see text) with size ratio~$\alpha$.  Whereas
the autocorrelation time per particle (expressed in $\mu$s of CPU time per
particle move) rapidly increases with size ratio, the GCA features only a weak
dependence on~$\alpha$.}
\label{fig:efficiency}
\end{center}
\end{figure}

The most striking feature of the generalized geometric cluster algorithm, apart
from the fact that it creates clusters that can be moved in a rejection-free
manner, is the speed at which it relaxes size-asymmetric mixtures. We
illustrate this here for a binary fluid mixture consisting of $N_{\rm s}$ small
and $N_{\rm l}$ large spherical particles with size ratio $\alpha \equiv
\sigma_{\rm l} / \sigma_{\rm s} > 1$. The particles are contained in a fixed
volume, at equal packing fractions $\eta_{\rm s} = \eta_{\rm l} = 0.1$. While
$N_{\rm l}=150$ is kept fixed, $N_{\rm s}$ increases from $1\,200$ to
$506\,250$ as $\alpha$ is varied from $2$ to~$15$.  Pairs of small particles
and pairs involving a large and a small particle act like hard spheres.
However, in order to prevent depletion-driven aggregation of the large
particles, they have a Yukawa repulsion,
\begin{equation}
\label{eq:yu}
U_{22}(r) = \left \{ \begin{array}{ll}
  +\infty                                              
  & \quad r \le \sigma_{\rm l}\\
  J \exp[-\kappa (r-\sigma_{\rm l})]/(r/ \sigma_{\rm l}) 
  & \quad r > \sigma_{\rm l}  \;,
  \end{array} \right.
\end{equation}
where $\beta J = 3.0$ and the screening length $\kappa^{-1} = \sigma_{\rm s}$.
In the simulation, the exponential tail is cut off at $3\sigma_{\rm l}$.  As a
measure of efficiency we consider the integrated autocorrelation time~$\tau$
obtained from the energy autocorrelation function~\cite{binder01},
\begin{equation}
\label{eq:correlation}
C(t) =
\frac{\langle{E(0)E(t)}\rangle -   {\langle{E(0)}\rangle}^2}%
      {\langle{E(0)^2}\rangle -  {\langle{E(0)}\rangle}^2} \;.
\end{equation}
For conventional MC calculations, $\tau$ rapidly increases with
increasing~$\alpha$, because the large particles tend to get trapped by the
small particles. Accordingly, an accurate estimate for $\tau$ could only be
obtained for $\alpha \leq 7$.  By contrast, the generalized GCA has an
autocorrelation time that only weakly depends on the size ratio, as illustrated
in Fig.~\ref{fig:efficiency}. At $\alpha=7$ the resulting efficiency gain
already amounts to more than three orders of magnitude.

\begin{figure}[p]
\begin{center}
\includegraphics[width=\figurewidth]{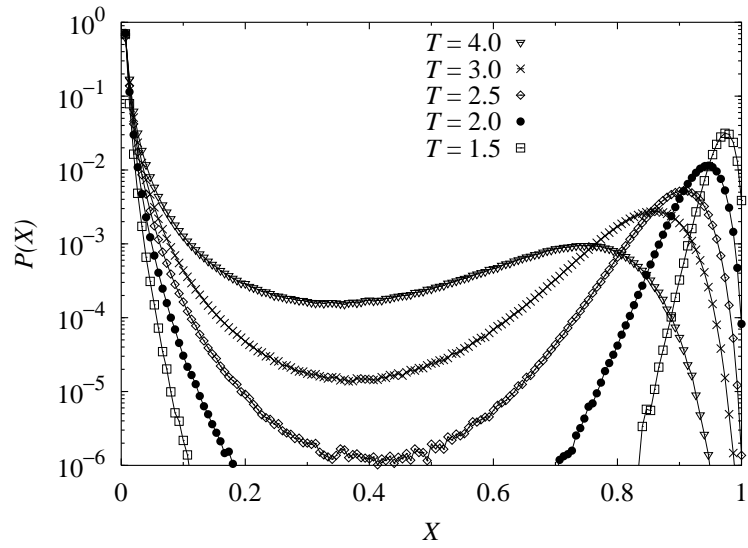}\\
(a)

\includegraphics[width=\figurewidth]{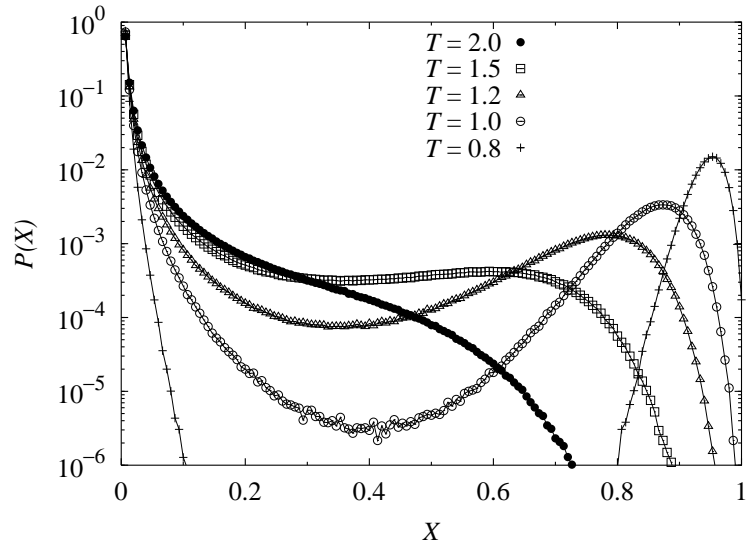}\\
(b)
\caption{Cluster size distributions as a function of relative cluster
size~$X$, for a monodisperse Lennard-Jones fluid. (a)~Reduced density $\rho^*
\equiv \rho\sigma^3 = 0.32$; (b) reduced density $\rho^*=0.16$. Identical
symbols compare to identical temperatures in both panels. All temperatures are
indicated in terms of $\varepsilon/k_{\rm B}$. See text for discussion.}
\label{fig:density}
\end{center}
\end{figure}

A crucial limitation of the generalized GCA is that each cluster must only
occupy a fraction of the entire system. As observed by Dress and
Krauth~\cite{dress95}, the entire system typically will be occupied by a single
cluster once the percolation threshold is reached in the combined system
containing a configuration and its point-reflected counterpart. For the
original system this leads, in three dimensions, to a practical upper limit in
the packing fraction around~$0.23$--$0.25$. This number will vary as a function
of size asymmetry and, if additional pair interactions are present,
temperature.  It is therefore instructive to study the cluster-size
distribution as a function of reduced density~$\rho^*$ and temperature~$T$.
Figure~\ref{fig:density} illustrates the cluster-size distributions as obtained
by means of the multiple-cluster GCA for a regular (one-component)
Lennard-Jones fluid. In the top panel, $\rho^*=0.32 \approx \rho^*_{\rm c}$.
Already at temperatures that are far above the critical temperature~$T_{\rm c}
\approx 1.19$, the cluster-size distribution starts to tend toward a bimodal
form, indicative of the formation of large clusters. In the vicinity of the
critical temperature, the average cluster size has become very large. For
comparison, the bottom panel displays the cluster-size distributions at a twice
smaller density, $\rho^*=0.16$. In this case the bimodal shape does not appear
until close to~$T_{\rm c}$.

\begin{figure}
\begin{center}
\includegraphics[width=\figurewidth]{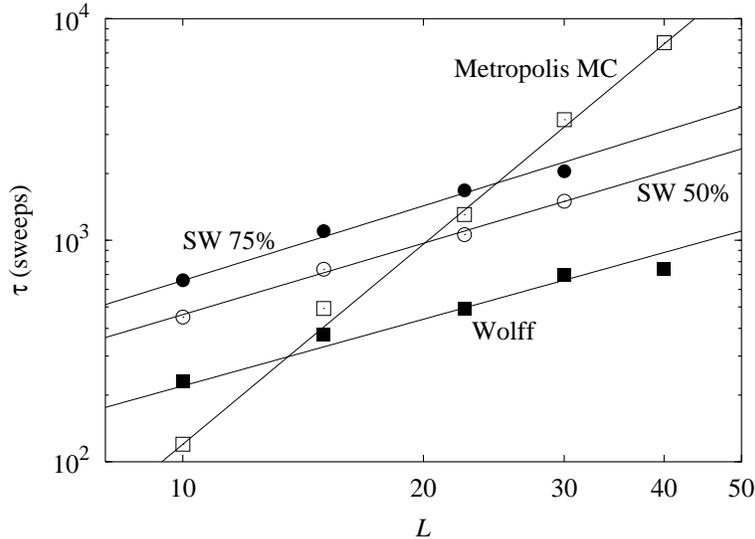}
\caption{Energy autocorrelation time $\tau$ as a function of linear
system size for a critical Lennard-Jones fluid, in units of particle sweeps,
for three different Monte Carlo algorithms: Local moves (``Metropolis MC'');
GCA with Swendsen--Wang type cluster decomposition and probability 0.50 (``SW
50\%'') and 0.75 (``SW 75\%'') of moving each cluster; single-cluster GCA
(``Wolff'').}
\label{fig:critdyn}
\end{center}
\end{figure}

The properties suggest that the generalized GCA, at least for the one-component
Lennard-Jones fluid, will not suppress critical slowing -- the primary
advantage of lattice cluster algorithms. For the off-lattice GCA, this property
is less essential, because of the speed-up it delivers for the simulation of
size-asymmetric fluids over a wide range of temperatures and packing fractions.
Nevertheless, we have investigated the integrated autocorrelation time for the
energy at the critical point, as a function of linear system size.  In
Fig.~\ref{fig:critdyn} these times are collected for three algorithms.
(1)~Conventional local-update Metropolis algorithm; (2)~Wolff version of the
GCA; (3a)~SW version of the GCA, in which each cluster is reflected with a
probability~$0.50$; (3b)~SW version of the GCA, in which each cluster is
reflected with a probability~$0.75$. Just as for spin models, the
single-cluster version outperforms the Swendsen--Wang type cluster
decompositions.  However, all variants of the GCA exhibit the same power-law
behavior, which outperforms the Metropolis algorithm by a factor $\sim
L^{2.1}$. It is important to emphasize that this acceleration may be due to the
suppression of the hydrodynamic slowing down~\cite{hohenberg77}
caused by the conservation of the density (which may couple to the energy
correlations~\cite{geomc}). Another striking point is that already for moderate
system sizes the generalized GCA outperforms the Metropolis algorithm, despite
the time-consuming construction of large clusters [cf.\ 
Fig.~\ref{fig:density}(a)] which only lead to small configurational changes.

\begin{figure}
\begin{center}
\includegraphics[width=\figurewidth]{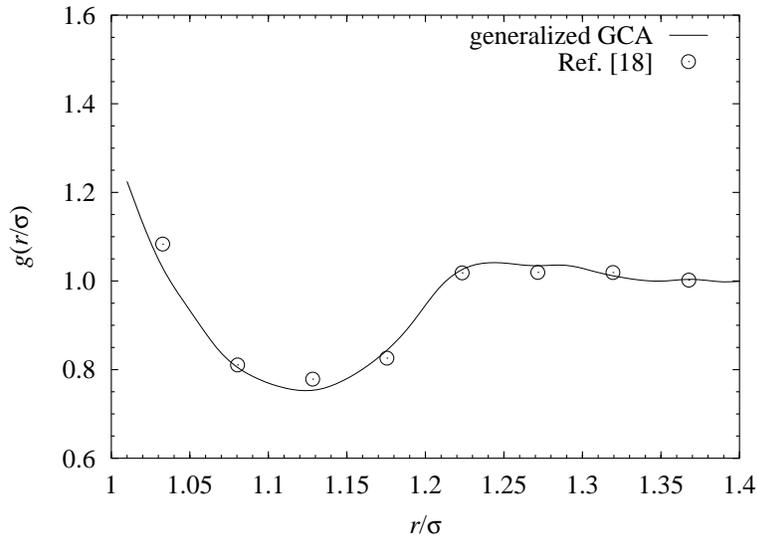}
\caption{Pair correlation function of dilute colloidal particles
(diameter~$\sigma$) in an environment of smaller particles
(diameter~$\sigma/5$, packing fraction~$\eta=0.116$) that experience a
Yukawa-type attractive interaction with the colloids. The symbols represent
data obtained by means of the hard-core GCA~\protect\cite{malherbe99}; the
solid line was obtained from the generalized GCA.}
\label{fig:amokrane}
\end{center}
\end{figure}

\section{Illustration}

In order to illustrate the capabilities of the generalized GCA, we have
computed the pair correlation function of dilute colloidal particles (packing
fraction $\eta_{\rm l}=0.001$) in an environment of particles with a five times
smaller diameter (packing fraction $\eta_{\rm s}=0.116$). Large and small
particles act as hard spheres, but unlike pairs (i.e., large--small) experience
a Yukawa attraction which promotes the accumulation of small particles around
the colloids. This system has been studied in Ref.~\cite{malherbe99} by means
of the original hard-core GCA, in which clusters are moved according to an
acceptance criterion, as proposed in Ref.~\cite{dress95}. This potentially
greatly deteriorates performance, as entire clusters will be constructed that
are subsequently rejected. Indeed, the authors report~\cite{malherbe99} that
the colloid pair correlation function $g(r)$ had to be obtained via numerical
differentiation of the integrated pair correlation function, rather than
through direct sampling. The generalized GCA can handle this system without
complication, as demonstrated in Fig.~\ref{fig:amokrane}. In order to obtain
data comparable to the solid curve in this figure, the authors of
Ref.~\cite{malherbe99} utilized a polynomial fit to the integrated pair
correlation function, which potentially leads to ambiguities.

\section{Conclusion and Outlook}

We have discussed the generalized geometric cluster algorithm, a Monte Carlo
method for the simulation of fluids by means of geometric operations. This
algorithm creates nonlocal multiple-particle moves that are capable of rapidly
decorrelating fluid configurations that contain particles of widely different
sizes. The multiple-particle moves are constructed in such a way that the
typical decrease in acceptance rate is avoided; every proposed move is accepted
without violating detailed balance. It is anticipated that this algorithm will
find widespread application, in particular in the simulation of complex fluids
and suspensions in which the solvent is modeled as an implicit background.
Potential generalizations include the treatment of particles with internal
degrees of freedom, such as nonspherical particles, and other geometries, such
as layered systems.

\section*{Acknowledgements}
This material is based upon work supported by the National Science Foundation
under grants nos.\ DMR-0346914 and CTS-0120978.


\end{document}